# A chameleon helioscope

Keith Baker [1], Axel Lindner [2], Amol Upadhye [3], Konstantin Zioutas [4]

[1] Physics Department, University of Yale, New Haven, USA;  oliver.baker@yale.edu

[2] DESY, Notkestraße 85, D-22607 Hamburg, Germany;  axel.lindner@desy.de

[3] Argonne National Laboratory, 9700 S. Cass Ave. Lemont, IL 60439, USA; aupadhye@hep.anl.gov

[4] University of Patras, GR 26504 Patras, Greece;  zioutas@cern.ch

**Abstract:**
Chameleon particles, which could explain dark energy, are in many ways similar to axions, suggesting that an axion helioscope can be used for chameleon detection. The distinguishing property of chameleon particles is that, unlike Standard Model particles, their effective masses depend upon the ambient matter–energy density. The associated total internal reflection of chameleons up to keV energies by a dense layer of material, which would occur at grazing incidence on the mirrors of an X-ray telescope, lead to new experimental techniques for detecting such particles.  In this short note we discuss when this total internal reflection can happen and how it can be implemented in existing or future state-of-the-art chameleon telescopes. Solar Chameleons would be emitted mainly with energies below a few keV suggesting the X-ray telescope as the basic component in chameleon telescopy. The implementation of this idea is straightforward, but it deserves further scrutiny. It seems promising to prepare and run a dark energy particle candidate detection experiment combining existing equipment.  For example, large volumes and strong solenoid magnetic fields, which are not appropriate for solar axion investigations, are attractive from the point of view of chameleon telescopy.

## 1. Introduction

Chameleons are particle candidates for the ubiquitous and mysterious dark energy in the Universe [1]. Chameleon dark energy is a canonical scalar field which couples conformally to matter and may also couple to photons. The interplay between this matter coupling and a nonlinear self-interaction gives the chameleon field a large mass in high-density regions of space, allowing it to evade stringent local constraints on fifth forces. We choose a potential $V(\phi)$ of runaway form [2], resulting in the effective potential

$$V_{\text{eff}}(\phi) = \Lambda^4 \left(1 + \frac{\Lambda^n}{\phi^n}\right) + \exp\left(\frac{\beta_m \phi}{M_{\text{Pl}}}\right)\rho_m + \frac{1}{4}\exp\left(\frac{\beta_\gamma \phi}{M_{\text{Pl}}}\right) F_{\mu\nu} F^{\mu\nu} \quad (1)$$

where $\Lambda = 2.4 \cdot 10^{-3}$ eV is the dark energy scale, $\rho_m$ is the matter density, and $F^{\mu\nu}$ is the electromagnetic field strength. In a constant-density bulk, the background field value $\phi_{min}$ is the minimum of $V_{\text{eff}}$, and the effective mass $m(\phi)$ is related to the curvature of $V_{\text{eff}}$:

$$\phi_{\text{min}}(\rho_m) = \Lambda \left(\frac{n M_{\text{Pl}} \Lambda^3}{\beta_m \rho_m}\right)^{\frac{1}{n+1}} \quad (2)$$

$$m(\phi)^2 = V''_{\text{eff}}(\phi) = \frac{n(n+1)\Lambda^{n+4}}{\phi^{n+2}} \quad (3)$$

Here we have made the approximations $\phi \ll M_{\text{Pl}}/\beta_m$, $M_{\text{Pl}}/\beta_\gamma$ appropriate to all cases of interest.
It has been suggested recently that chameleon particles can be produced inside the macroscopic inner solar magnetic fields near the tachocline, i.e., a shell at about $0.7 \cdot R_{\text{solar}}$ [3]. In fact, a manifestation of the widely predicted 50-100 Tesla inner solar fields at the tachocline are the magnetic fields at the solar surface, and in particular the strong ones (up to a few kGauss) at the sunspots. The underlying production mechanism for the solar chameleons is the celebrated Primakoff-effect, as the thermal photons propagate outwards in a random walk inside magnetized regions. The estimated spectral shape of the solar chameleon spectrum, which should escape from the tachocline region, is a broad distribution peaking at ~600 eV. Interestingly, the escaping luminosity in an as yet invisible channel like that of chameleons can be as much as ~10% of the total solar luminosity ($L_{\text{tot}}=3.8 \cdot 10^{33}$ erg/s). This is a relatively large fraction, without coming though in conflict with the evolution of the Sun over its ~4.5 Gyears [4].
Since there are very few experiments aiming to detect dark energy candidate particles, any new approach or improvement towards a direct detection and particle identification of such exotica is of utmost importance, and therefore it deserves further consideration. With this short note we point at the peculiar dynamical behavior of chameleons when they are approaching the surface of a dense object as a major ingredient for any chameleon detector. In the case of solar chameleons, state-of-the-art X-ray telescopes can be utilized straightforwardly in the CERN Axion Solar Telescope (CAST), but also in other types of telescopes. The special properties of solar chameleons can prescribe the design of novel telescopes which avoid the usual severe requirements on the mirrors.

## 2. Focusing chameleons

It will be shown in the next section that a dense material is only transparent to chameleon particles above a critical energy and below a maximum angle of incidence. Consequently, chameleons below a critical energy and grazing angle will undergo a total reflection; chameleons bounce at grazing incidence. This work is inspired by this unique property of chameleon particles. Existing telescopes capable of detecting soft X-rays could also focus the ≈ 0.6 keV chameleons produced in the Sun or the laboratory [3]. Reflection at grazing incidence can be used to direct chameleon particles toward a focal point. A magnetic field, such as those in existing magnetic telescopes [5] or in a high-field solenoid magnet placed near the focal point, would allow those chameleon particles to oscillate into photons which could then be observed by an X-ray detector. The essential point is that existing technology can focus the chameleon dark energy particles themselves, and not just the photons which they produce after oscillation. This property of chameleon particles offers new ways to detect them and to distinguish them from other exotic particles. An experiment such as CAST [5], rather than just accepting particles which enter a narrow magnetic pipe, could use a focusing X-ray mirror system with a large collecting area to direct chameleon particles into the pipe, increasing the chameleon flux by orders of magnitude. The mirror system, such as the one in CAST as well as those in astrophysical X-ray telescopes, need not be in a vacuum and can therefore be placed outside of the magnet. As in the "chameleon-shining-through-a-wall" experiment proposed by [3], a beam dump placed at the intake of the magnetic pipe would ensure that no photons were entering the pipe. Alternatively, a dedicated chameleon helioscope could place a high-field solenoid magnet near the focus of the chameleon beam, with the magnetic field transverse to the particle direction. The large chameleon flux near the focus means that the magnetic field need not necessarily extend over a large spatial region. More generally, chameleon focusing allows a helioscope with a laterally small magnetic field size to have a large collecting area, implying that constraints on chameleon particles could be made substantially stronger than those on axions. Though a detailed discussion of helioscope configurations is beyond the scope of this paper, the focusing of dark energy particles is a promising method for enhancing the signal that a helioscope could expect to measure. For example, utilizing the entire CAST X-ray telescope or XMM/Newton, the effective solar chameleon flux in CAST can increase by as much as a factor of about 10 and 300, respectively.

## 3. The dynamical behavior of chameleons near interfaces

Consider the chameleon field outside a thick planar slab with density $\rho(x) = \rho_m$ for $x < 0$ and $\rho(x) = 0$ for $x \geq 0$. The static field profile and mass at $x \geq 0$ can be found exactly [6,7]:

$$\phi_S = (1+1/n)\phi_{\min}(\rho_m) \tag{4}$$

$$m_S^2 = n(n+1) \Lambda^{4+n}/\phi_S^{n+2} \tag{5}$$

$$\phi_0(x) = \phi_S \left(1 + \frac{(n+2)m_S x}{\sqrt{2n(n+1)}}\right)^{\frac{2}{n+2}} \tag{6}$$

$$m_0(x) = m_S\left(1 + \frac{(n+2)m_S x}{\sqrt{2n(n+1)}}\right)^{-1} \tag{7}$$

Here $\phi_S$ and $m_S$ are the field and mass on the surface $x=0$ of the planar slab. Meanwhile, at $x<0$, $\phi_0$ and $m_0$ approach $\phi_{\min}(\rho_m)$ and $m(\phi_{min}(\rho_m))$, respectively, over a length scale of order $m_S^{-1}$. Thus for practical purposes the slab may be considered "thick" if its thickness is much greater than $m_S^{-1}$.

The reflection of chameleon plane waves was studied in [7]; we summarize those results here. Perturbing about the static chameleon field, $\phi(\vec{r},t) = \phi_0(x) + \delta\varphi(\vec{r},t)$, we find the equation of motion

$$(\Box - m_0^2)\delta\phi = 0 \tag{8}$$

for the perturbation $\delta\phi$. This is essentially a statement of energy conservation for a chameleon particle. If the particle energy far from matter is $\omega$, then that particle will be excluded from the region of space where $m_0 > \omega$. Thus an incident particle will reflect from the matter, if $\omega < m(\phi_{min}(\rho_m))$. For a plane wave at normal incidence, we write $\delta\phi = e^{-i\omega t}\widetilde{\delta\phi}$. Then relation (8) reduces to an ordinary differential equation whose solution is $\widetilde{\delta\phi} \propto \sqrt{\omega x}J_\alpha(\omega x)$, where $\alpha = \frac{3n+2}{2n+4}$ and $J_\alpha$ is the regular Bessel function [7].

This can easily be generalized to nonzero incident angles [8]. Let the plane of reflection be the $xy$ plane. Since relations (6) and (8) are independent of $y$, the particle momentum $k_y$ in the $y$ direction is conserved. Thus we factor out the $y$-dependent part of the solution as well, writing $\delta\phi = e^{ik_y y - i\omega t}\tilde{\delta}\phi$. The solution is unchanged except that $\omega$ is replaced by $k_{x,\infty} = \left(\omega^2 - k_y^2\right)^{1/2}$, the momentum in the $x$ direction at large $x$. Furthermore, the particle will reflect, if $m_0 > k_{x,\infty}$ at some $x$. Thus even a chameleon particle with $\omega \gg m_s$ can reflect at grazing incidence, if $k_{x,\infty} < m(\phi_{min}(\rho_m))$.

We wish to determine whether a chameleon particle in space would be able to reach the Earth's surface. Treating the atmosphere as a planar slab of density $\rho_m = 1.2 \cdot 10^{-3}$ g/cm$^3$, we can compute the chameleon mass $m(\phi_{min}(\rho_m))$, which is the minimum value that $k_{x,\infty}$ must take, if the particle is to be transmitted through the atmosphere; here the $x$ direction is normal to the Earth's surface. Figure 1 (left) shows this mass as a function of the chameleon model parameters. Given an incident angle $\theta = arccos(k_{x,\infty}/\omega)$, we find that $\omega > m(\phi_{min}(\rho_m))/\cos(\theta)$ is required for transmission. Figure 1 (right) shows this minimum transmission energy as a function of incident angle for various models.

Next, we show that high-energy chameleon particles transmitted through the atmosphere can be focused by the mirrors of an X-ray telescope, which reflect particles at grazing incidence. Let $\varepsilon = 90°-\theta$ be the grazing angle. Then the reflection condition $k_{x,\infty} < m(\phi_{min}(\rho_m))$ becomes $\omega < m(\phi_{min}(\rho_m))/\sin(\varepsilon)$. Figure 2 shows this maximum energy which can be focused as a function of incident angle for several chameleon models. Since $\varepsilon$ can be quite small, chameleon particles with energies much greater than $m(\phi_{min}(\rho_m))$ can be reflected. As a concrete example, consider the model with n=4 and $\beta_m = 10^6$ incident on a material with $\rho_m = 10$ g/cm$^3$. The effective mass inside the material is 22 eV; at normal incidence, a chameleon particle with more energy than this will be transmitted through the material. At a grazing angle of, say, 30', particles with energies up to 2.5 keV will reflect from the surface. This means that chameleons with energies in the sub-keV range, such as those expected to be emitted from the Sun's magnetized tachocline [3], can be totally reflected by the mirrors of X-ray telescopes. This resembles the reflection of

X-rays at small grazing angles, even though the underlying physics is quite different. X-ray reflection is due to coherent Compton scattering of the photons with the mirror electrons. In contrast, chameleon particles reflect specularly from the background chameleon field extending beyond the mirror surface.

The classical turning point for a bouncing chameleon plane wave is at a distance $\sim k_{x,\infty}^{-1} > m(\phi_{min}(\rho_m))^{-1}$ outside the mirror. Thus the reflection will not be affected by imperfections in the mirror surface on scales much smaller than this. From the example discussed above, $m(\phi_{min}(\rho_m))^{-1} = (22 \text{ eV})^{-1} \sim 10$ nm. As another example, consider a $\omega = 600$ eV chameleon with a grazing angle of 30′, implying $k_{x,\infty} = 5.2$ eV and $k_{x,\infty}^{-1} = 38$ nm. Once again we see that the classical turning point is far enough from the surface for the reflection to be affected by Angstrom-scale surface imperfections. Thus a dedicated chameleon helioscope will not need the highly polished mirrors which are required for conventional X-ray telescopes, allowing it to be built at a substantially lower cost.

Of course, X-ray telescopes with well-polished mirrors exist. Such an instrument can be an integral part of a chameleon helioscope or telescope. X-ray telescopes can thus guide the chameleon "beam" from the Sun, or other sources in the sky, into a small magnetic pipe with the strongest possible magnetic field, giving rise to the emission of more X-rays from converted chameleons. A chameleon helioscope should be cheaper to construct than an axion helioscope!

In this short note, actually we are not aiming to elaborate every configuration detail of all the potential types of new chameleon detector systems, which depends on the actual magnetic field to be used. We simply point out that the expected focusing and imaging of dark sector particles opens new opportunities for this very young research field.

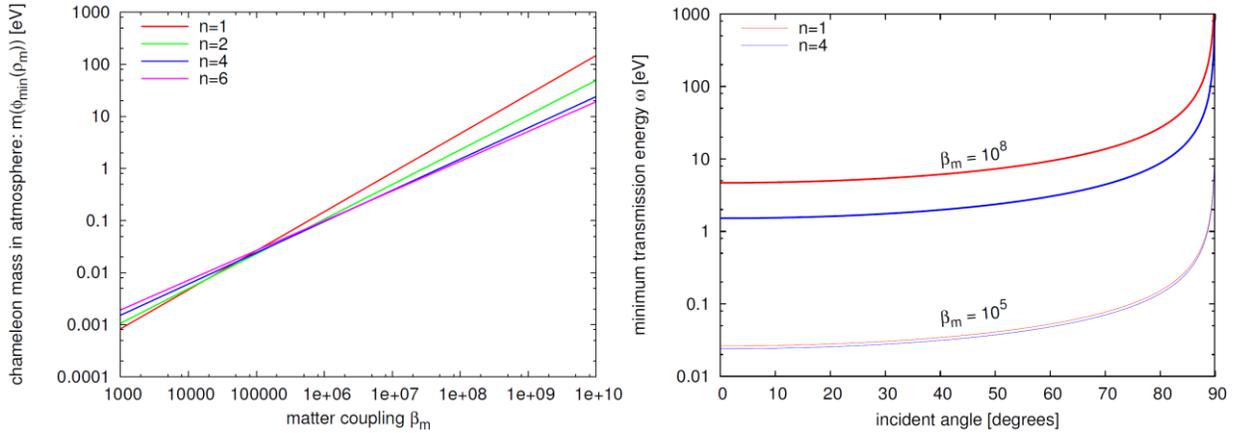

**Figure 1** Minimum chameleon energy ω required for transmission through the atmosphere to the Earth's surface. The atmosphere has at sea level a density of $\rho_m \approx 1.2 \cdot 10^{-3}$ g/cm$^3$. (*left*) The minimum energy at normal incidence is the chameleon mass in the atmosphere, shown as a function of $\beta_m$ and n. (*right*) Chameleons at nonzero incident angles θ require greater energies for transmission, i.e., when the chameleons hit the plane of the denser surface less and less perpendicularly, more and more energetic chameleons can be reflected.

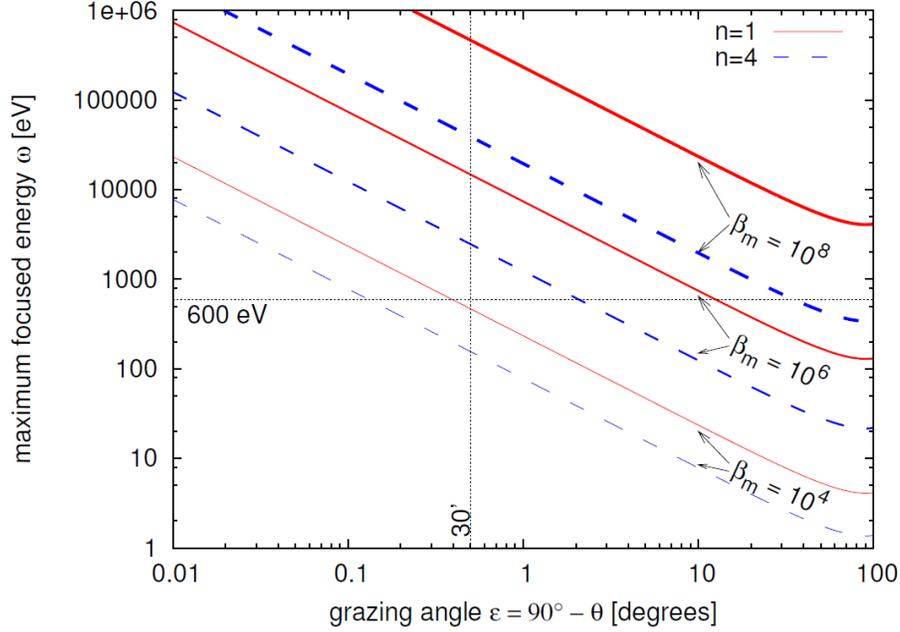

**Figure 2** Maximum energy at which a chameleon particle can be focused by an X-ray mirror with density $10 \, \text{g/cm}^3$ ($\approx$ the density of a Ni-coated X-ray telescope) and grazing angle $\varepsilon$, for several different chameleon models. The dotted horizontal and vertical lines illustrate one example of a 600 eV chameleon incident on a mirror of focusing angle 30', which is, for example, equal to the field-of-view of XMM/Newton. The chameleon will be focused by this mirror if n=4 and $\beta_m=10^6$, but will pass through the mirror if n=1 and $\beta_m=10^4$.

## 4. Conclusions

A theoretically motivated particle candidate for the mysterious and ubiquitous dark energy in the universe, the chameleon has properties quite different from known, Standard Model particles. These allow entirely new avenues for the detection of chameleon dark energy which would require simple modifications to existing telescopes. For example, the chameleon mass is not universal; chameleon particles acquire an effective mass which grows with the ambient matter density. Because of this property they can be reflected by dense matter, including ordinary laboratory materials as well as astrophysical objects. In this short note we have outlined a general framework for detecting or constraining such particles coming from the Sun or from outer space. This makes nearly all existing X-ray telescopes, particularly those in orbit, of potential interest.

More specifically, if chameleons come from the inner Sun, existing X-ray telescopes are best suited for their detection and identification. So far CAST uses an X-ray telescope for focusing / imaging purposes of the secondary X-rays expected to emerge from the interaction of solar axions with the intervening magnetic field. In the case of solar chameleons, the whole X-ray telescope can focus a much larger flux of incident chameleons, directing them into the magnetic field region towards the field end, where they can be converted to (soft) X-rays following the inverse Primakoff - effect inside the magnetic field (similar to axions). Under certain conditions, this allows to utilize the entire area of the largest X-rays telescopes, increasing thus the flux of such particles to be "seen" by the detector.

In fact, it appears to be an advantage in building a dedicated chameleon telescope compared to an axion helioscope, which requires a very large and strong magnetic field to allow a sufficient number of axions to convert into X-rays, which then are focused onto a detector. For chameleons, one could first focus the chameleons with a large telescope (even vacuum is not required!) and then guide the chameleon beam into small sized magnetic pipes utilizing thus a very strong magnetic field, which will allow to produce more excess X-rays. This design should be much cheaper to construct than an axion helioscope!

At present, the field of X-ray astronomy is a quite active area of research also for astroparticle physics purposes, which is best illustrated by the construction of powerful and large area X-rays telescopes for the NuSTAR mission, which can be used for chameleon searches too. The proposed design for a new axion telescope IAXO [9], might profit by implementing the suggested scheme. The same reasoning applies equally well to other telescopes too, whose feasibility is currently under investigation. For example, large volume and strong solenoid magnetic fields, which are not appropriate for solar axion investigations, are attractive within the reasoning of this work.

Finally, it is worth noticing that chameleons should be seen as a generic example of such particle candidates of the ubiquitous dark energy sector in the Universe. In any case, strong and large volume magnetic fields seem to be a basic ingredient for the field of astroparticle physics. We have suggested in this work the working principle of focusing solar chameleons by using existing equipment in astrophysics like the X-ray telescopes, or other type of telescopes. We recall that solar chameleons are particle candidates from the dark energy sector, and the new idea is that one can focus or even guide them, which otherwise only gravitational lensing can do, and this is the novelty of this work.